\documentclass[preprint,review,12pt]{elsarticle}

\usepackage{amssymb}
\usepackage{amsmath}
\usepackage{esint}
\usepackage{amssymb}
\usepackage{amsmath}
\usepackage{caption} %改变图表标题
\usepackage{booktabs} %调整表格线与上下内容的间隔
\usepackage{longtable}%调用跨页表格
\usepackage{multirow} %多行合并
\usepackage{array} %调用公式宏包的命令应放在调用定理宏包命令之前，也能控制表格
\usepackage{graphicx}
\usepackage{subcaption}
\usepackage{float}
\usepackage{hyperref}

\journal{Journal of Computational Physics}

\begin{document}

\begin{frontmatter}



\title{A Flow-Based Hybrid Approach for Kinetic Plasma Simulations: Bridging Direct Vlasov and Particle Methods}


\author[label1]{Bowen Zhu} % First author
\author[label1]{Jian Wu\corref{cor1}} 
\ead{jxjawj@mail.xjtu.edu.cn}
\author[label1]{Yuanbo Lu}
\affiliation[label1]{organization={School of Electrical Engineering},% Department and Organization
            addressline={Xi'an Jiaotong University}, 
            city={Xi'an},
            state={Shaanxi},
            country={China}}

%% Corresponding author details
\cortext[cor1]{Corresponding author}
% \ead{jxjawj@mail.xjtu.edu.cn} % Replace with the corresponding author's email

%% Abstract
\begin{abstract}
We present a novel flow-based kinetic approach, inspired by continuous normalizing flows, for plasma simulation that unifies the complementary strengths of direct Vlasov solvers and particle-based methods. By tracking the distribution function along the characteristic curves defined by the Newton--Lorentz equations, our method directly computes $f(\mathbf{z}(t))$ at selected points in phase space without reliance on Monte Carlo sampling.

We employ a scatter-point integration scheme using smoothing kernels reminiscent of Smoothed Particle Hydrodynamics (SPH), to calculate field quantities and moments, achieving higher accuracy with far fewer markers compared to Particle-in-Cell (PIC) methods.

Unlike PIC, our approach supports strategic marker placement and dynamic refinement in regions of interest, thus reducing sampling noise and computational overhead. This capability is particularly advantageous in high-density plasmas, where PIC’s particle requirements can be prohibitive. In addition, the method naturally accommodates collisional effects via an augmented phase-space flow description  ensuring robust handling of both collisionless and collisional plasmas.

Our simulations of Landau damping, two-stream instability, and collisional relaxation demonstrate reduced noise, accurate phase-space resolution with significantly fewer markers, and robust energy conservation. Moreover, the independent characteristic curves and local scatter integration are highly amenable to GPU acceleration, enabling efficient large-scale simulations.

Overall, this flow-based framework offers a powerful, flexible, and computationally efficient alternative to traditional particle methods for kinetic plasma dynamics, with potential applications spanning inertial confinement, Zpinch, and other complex kinetic systems.

\end{abstract}

\begin{graphicalabstract}
\end{graphicalabstract}

\begin{highlights}
\item Novel Hybrid Methodology:

The paper presents a new flow-based approach that successfully bridges direct Vlasov solvers and particle methods for kinetic plasma simulations. By tracking distribution functions according to the theory of continuous normalizing flow, it combines the benefits of both methods.

\item Superior Convergence Performance

 The method demonstrates significantly improved accuracy with far fewer computational markers compared to traditional Particle-in-Cell (PIC) approaches. Specifically, it achieves comparable results using about 100x fewer markers than PIC while maintaining good energy conservation properties.

\item Flexible Adaptive Resolution

Flexible Adaptive Resolution: The method enables strategic marker placement and dynamic refinement in regions of interest, providing enhanced resolution where needed without the sampling noise issues inherent in PIC methods.

\item Unified Treatment of Physics

The approach naturally accommodates both collisionless and collisional plasma dynamics within a single framework through an augmented phase-space flow description.

% , handling phenomena like Landau damping, two-stream instability, and collisional relaxation.

\item High Density Plasma Advantage

The method shows particular promise for high-density plasma simulations where traditional PIC methods become computationally prohibitive.
\end{highlights}

%% Keywords

\end{frontmatter}



\section{Introduction}
Plasma simulations pose significant challenges in computational physics, spanning scales from microscopic particle interactions to macroscopic collective phenomena. At the kinetic level, plasma evolution is described by the Vlasov-Maxwell system for collisionless plasmas or the Fokker-Planck system for collisional plasmas, capturing the distribution function \(f(\mathbf{x}, \mathbf{v}, t)\) in a six-dimensional phase space \cite{boyd2003physics}. Although these equations comprehensively represent plasma behavior, their high dimensionality severely impedes direct numerical solutions .

Numerical methods for kinetic plasma simulation have historically taken two primary routes. On one hand, direct Vlasov solvers work on phase-space grids and can achieve high accuracy but suffer from the curse of dimensionality \cite{filbet2001conservative}. On the other hand, particle-in-cell (PIC) methods \cite{birdsall2018plasma} exploit the equivalence between the Vlasov equation and the Newton–Lorentz dynamics of individual particles, approximating \(f\) with ensembles of computational particles. Despite their success, PIC methods rely on Monte Carlo integration, which converges as \(\mathcal{O}(N^{-1/2})\) in the number of particles \cite{robert1999monte}. In practice, achieving high fidelity often demands prohibitively large particle counts, restricting PIC to relatively low-density plasmas or shorter simulation times. While refinements such as the \(\delta f\) method focus computational resources on small deviations from an equilibrium state to reduce sampling noise \cite{parker1993fully}, they remain fundamentally tied to PIC and often require near-equilibrium assumptions.

Concurrently, recent progress in machine learning, especially in generative modeling, has led to the development of continuous normalizing flows (CNFs) \cite{chen2018neural}. In CNFs, neural ordinary differential equations evolve probability densities in continuous time \cite{grathwohl2018ffjord}, mirroring the way the Vlasov equation evolves a phase-space distribution. This mathematical similarity suggests a compelling bridge between traditional Vlasov solvers and particle-based methods.

The key insight of this work is recognizing that the instantaneous change-of-variables formula \cite{grathwohl2018ffjord} from CNF theory provides a natural framework for tracking the distribution function along characteristic curves. Coupling this to the Newton–Lorentz equations governing particle trajectories yields an explicit computation of \(f(\mathbf{z}(t))\) at arbitrary points \(\mathbf{z}(t) = [\mathbf{x}(t), \mathbf{v}(t)]\) in phase space. 

Notably, these points \(\mathbf{z}\) evolve over time but need not be interpreted as physical particles; rather, they serve as convenient phase-space markers. Moreover, this approach can naturally incorporate collisional effects by augmenting the phase-space flow to include collision operators.

In this paper, we propose a new numerical method that blends elements from direct Vlasov solvers, particle approaches, and normalizing flows. By tracking the distribution function along trajectories, we solve the Vlasov equation in a way that avoids both phase-space gridding and large sampling noise. For field calculations, we introduce a scatter-point integration scheme inspired by Smoothed Particle Hydrodynamics (SPH) \cite{monaghan1992smoothed}. Unlike PIC-based Monte Carlo methods, this scheme supports flexible marker placement, including spawning markers adaptively during the simulation. Such flexibility not only facilitates targeted resolution in regions of interest but also offers superior convergence properties.

Our approach provides several key advantages:
\begin{enumerate}
    \item \textbf{Increased information content per computational marker:} Direct tracking of the distribution function allows each marker to carry a piecewise representation of \(f\).
    \item \textbf{Freedom from Monte Carlo sampling constraints:} Eliminating random sampling enables more efficient resource allocation.
    \item \textbf{Straightforward incorporation of additional physics:} Collision models and other effects can be included by modifying the phase-space flow.
    \item \textbf{Adaptive refinement:} New markers can be introduced during the simulation, focusing computational effort where needed.
    % \item \textbf{Compatibility with GPU acceleration:} The mathematical structure is amenable to modern parallel architectures.
\end{enumerate}

We demonstrate the effectiveness of the proposed method through a series of numerical experiments, reproducing classical plasma phenomena such as Landau damping, two-stream instability, and collisional dynamics. 

The paper is structured as follows: Section 2 reviews kinetic plasma simulation and continuous normalizing flows, connecting the Vlasov equation to normalizing flows. Section 3 details the numerical implementation, covering time advancement, SPH-inspired integration, adaptive refinement, and convergence criteria. Section 4 validates the approach against theoretical results. Section 5 analyzes computational costs, and Section 6 presents conclusions and future directions.

\section{Theoretical Framework}

\subsection{Equivalence of the Vlasov Equation and Newton--Lorentz Particle Dynamics}
\label{sec:equivalence_vlasov_particle}
The collisionless Vlasov equation governs the evolution of a distribution function \( f(\mathbf{x}, \mathbf{v}, t) \) in phase space, under self-consistent electromagnetic fields. For a nonrelativistic, single-species plasma, it takes the form~\cite{chen1984introduction}:
\begin{equation}
    \frac{\partial f}{\partial t} 
    + \mathbf{v} \cdot \nabla_{\mathbf{x}} f
    + \frac{q}{m} \bigl(\mathbf{E} + \mathbf{v} \times \mathbf{B}\bigr) 
      \cdot \nabla_{\mathbf{v}} f 
    = 0,
    \label{eq:vlasov}
\end{equation}
where \( q \) and \( m \) denote the charge and mass of the particles, respectively, and \( \mathbf{E} \) and \( \mathbf{B} \) are the electric and magnetic fields. This equation is typically coupled to Maxwell's equations to form the self-consistent Vlasov--Maxwell system.

The equivalence between the Vlasov equation and the Newton--Lorentz particle dynamics can be shown via the method of characteristics~\cite{nicholson1983introduction}. Consider the phase-space trajectory \(\mathbf{z}(t) = [\mathbf{x}(t), \mathbf{v}(t)]\) governed by
\begin{subequations}
\label{eq:newton_lorentz}
\begin{align}
    \frac{d\mathbf{x}}{dt} &= \mathbf{v}, \\
    \frac{d\mathbf{v}}{dt} &= \frac{q}{m} \bigl[\mathbf{E}(\mathbf{x}, t) 
        + \mathbf{v} \times \mathbf{B}(\mathbf{x}, t)\bigr].
\end{align}
\end{subequations}
The total time derivative of \( f \) along this trajectory is
\begin{equation}
    \frac{d f}{dt} \;=\; \frac{\partial f}{\partial t} 
    + \frac{d\mathbf{x}}{dt} \cdot \nabla_{\mathbf{x}} f 
    + \frac{d\mathbf{v}}{dt} \cdot \nabla_{\mathbf{v}} f.
\end{equation}
Substituting Eqs.~\eqref{eq:newton_lorentz} into this derivative reproduces Eq.~\eqref{eq:vlasov}, thereby identifying the characteristic curves of the Vlasov equation with Newton--Lorentz trajectories.

This equivalence underpins particle-based simulation techniques such as the Particle-In-Cell (PIC) method. In PIC, a finite number of macro-particles sample the phase space, evolving under Eqs.~\eqref{eq:newton_lorentz}. The resulting charge and current densities are used as sources in Maxwell's equations, whose solutions update the electromagnetic fields that act back on the particles. Thus, the coupled Vlasov--Maxwell system is effectively solved iteratively.

Although Eq.~\eqref{eq:vlasov} applies strictly to collisionless plasmas, additional physics can be incorporated. For instance, collisional effects enter via a collision operator, \( C[f] \), on the right-hand side:
\begin{equation}
\frac{\partial f}{\partial t}
+ \mathbf{v} \cdot \nabla_{\mathbf{x}} f
+ \frac{q}{m} \bigl(\mathbf{E} + \mathbf{v} \times \mathbf{B}\bigr)
\cdot \nabla_{\mathbf{v}} f
= C[f].
\label{eq:vlasov_collision}
\end{equation}
In particle-based methods, such as PIC, collisions are often treated through Monte Carlo (MC) models that stochastically sample pairwise interactions within each cell \cite{takizuka1977binary}. When a collision event occurs, the post-collision velocities are updated according to scattering laws and cross sections. Additional physics such as: radiation reaction, on the other hand, can be described by adding a radiative damping term to Eq.~\eqref{eq:newton_lorentz}\textsubscript{b}. A commonly used form is the Landau--Lifshitz force, which accounts for self-force effects arising from the radiation emitted by accelerating charges~\cite{landau2013classical}. Other implementations of radiation processes can similarly be introduced in the particle equations of motion by adding appropriate force terms.

Our objective is to combine the strengths of both direct Vlasov methods and particle-based PIC approaches. To this end, we harness the concept of \emph{continuous normalizing flows} to bridge these numerical frameworks. The details of our approach, including implementation and performance comparisons, are presented in the following sections.

\subsection{Analogy with Continuous Normalizing Flows}
\label{sec:cnf_analogy}

Normalizing flows provide a powerful approach to transform an initial probability distribution into a target distribution via a sequence of invertible transformations. In particular, \emph{continuous normalizing flows} (CNFs)~\cite{chen2018neural} represent this sequence by a time-dependent ordinary differential equation (ODE). Specifically, let $\mathbf{z}(t)\in\mathbb{R}^{d}$ be a sample evolving under 
\begin{equation}
    \frac{d\mathbf{z}(t)}{dt} = \mathbf{F}\bigl(\mathbf{z}(t), t;\boldsymbol{\theta}\bigr),
    \quad \mathbf{z}(0)=\mathbf{z}_0,
    \label{eq:cnf_ode}
\end{equation}
where $\mathbf{F}$ is typically a neural network with parameters $\boldsymbol{\theta}$. If $p(\mathbf{z}(t),t)$ denotes the probability density of $\mathbf{z}(t)$, then the \emph{instantaneous change of variables} formula~\cite{grathwohl2018ffjord} asserts:

\begin{subequations}
\begin{align}
    \log p\bigl(\mathbf{z}(t), t\bigr)
    = \log p_0(\mathbf{z}_0) 
      - \int_{0}^{t} \nabla_{\mathbf{z}(\tau)} 
        \cdot \mathbf{F}\bigl(\mathbf{z}(\tau), \tau;\boldsymbol{\theta}\bigr)
        \,d\tau, \quad or \\
    d \log p\bigl(\mathbf{z}(t), t\bigr)/dt
    =  
      -  \nabla_{\mathbf{z}(\tau)} 
        \cdot \mathbf{F}\bigl(\mathbf{z}(\tau), \tau;\boldsymbol{\theta}\bigr)
        \,d\tau,
\label{eq:cnf_log_density}
\end{align}
\end{subequations}

where $p_0(\mathbf{z}_0)$ is the initial density at $t=0$. This mechanism accounts for expansion or contraction of the volume in $\mathbf{z}$-space via $\nabla\cdot \mathbf{F}$.

In the kinetic description of plasmas, we consider a phase-space vector 
\[
    \mathbf{z} = 
    \begin{pmatrix}
        \mathbf{x} \\
        \mathbf{v}
    \end{pmatrix},
\]
where $\mathbf{x}$ and $\mathbf{v}$ denote the position and velocity of a charged particle. The \emph{Vlasov equation} dictates how the distribution function $f(\mathbf{z},t)$ evolves in time; its characteristic form is governed by the Newton--Lorentz equations:
\begin{equation}
    \frac{d\mathbf{z}(t)}{dt} 
    = \mathbf{G}\bigl(\mathbf{z}(t), t; \mathbf{E}, \mathbf{B}\bigr)
    = 
    \begin{pmatrix}
        \mathbf{v}(t) \\[6pt]
        \frac{q}{m} 
        \bigl[\mathbf{E}(\mathbf{x}(t), t) + \mathbf{v}(t)\times \mathbf{B}(\mathbf{x}(t), t)\bigr]
    \end{pmatrix},
    \label{eq:phase_space_flow}
\end{equation}
where $q$ and $m$ are the particle charge and mass, respectively, and $\mathbf{E}$ and $\mathbf{B}$ are the self-consistent electric and magnetic fields. 

Comparing Eq.~\eqref{eq:phase_space_flow} with the CNF ODE Eq.~\eqref{eq:cnf_ode}, it is natural to ask how $f(\mathbf{z},t)$, viewed as a probability density in phase space, changes in time. If we were to \emph{naively} apply the same instantaneous change of variables formula Eq.~\eqref{eq:cnf_log_density} to $f(\mathbf{z},t)$, we would write
\begin{equation}
    \log f\bigl(\mathbf{z}(t), t\bigr)
    = \log f(\mathbf{z}_0,0)
      - \int_0^t \nabla_{\mathbf{z}(\tau)} \cdot 
        \mathbf{G}\bigl(\mathbf{z}(\tau), \tau;\mathbf{E}, \mathbf{B}\bigr)\,d\tau.
    \label{eq:log_density_phase_space}
\end{equation}

To compute 
\(
\nabla_{\mathbf{z}} \cdot \mathbf{G},
\)
we split \(\mathbf{z}\) into \((\mathbf{x},\mathbf{v})\) and write out the divergence in 6-dimensional phase space:

\[
\nabla_{\mathbf{z}} \cdot \mathbf{G}
=
\underbrace{\sum_{i=1}^{3}
\frac{\partial}{\partial x_i}\bigl[v_i\bigr]}_{\displaystyle 0}
\;+\;
\underbrace{\sum_{i=1}^{3}
\frac{\partial}{\partial v_i}
\biggl[
\dfrac{q}{m}\Bigl(E_i(\mathbf{x},t) + \bigl(\mathbf{v}\times\mathbf{B}(\mathbf{x},t)\bigr)_i\Bigr]
\biggr]}_{\displaystyle 0}.
\]
Since \(\mathbf{v}\) is treated as independent of \(\mathbf{x}\) in phase space, the partial derivatives 
\(\tfrac{\partial v_i}{\partial x_j}\) vanish. 
Similarly, 
\(\tfrac{\partial}{\partial v_i}\bigl[E_i(\mathbf{x},t)\bigr] = 0\) 
and
\[
\sum_{i=1}^3 \frac{\partial}{\partial v_i}\Bigl[\bigl(\mathbf{v}\times \mathbf{B}\bigr)_i\Bigr]
=
\sum_{i=1}^3 \frac{\partial}{\partial v_i} 
\bigl(\epsilon_{ijk}\,v_j\,B_k(\mathbf{x},t)\bigr)
=
\sum_{i=1}^3 \epsilon_{ijk}\,B_k(\mathbf{x},t)\,\delta_{j i}
= 
\sum_{i=1}^3 \epsilon_{i i k}\,B_k
= 0,
\]
where \(\epsilon_{ijk}\) is the Levi-Civita symbol and \(\delta_{ji}\) is the Kronecker delta. Hence,
\[
\nabla_{\mathbf{z}} \cdot \mathbf{G}
\;=\;
0.
\]

The result follows from Hamiltonian systems without collisional or dissipative effects; 
Liouville's theorem ensures \(\nabla_{\mathbf{z}} \cdot \mathbf{G} = 0\), indicating 
that the phase-space flow is incompressible and volume-preserving. Consequently,
\begin{equation}
    \nabla_{\mathbf{z}(\tau)} \cdot 
    \mathbf{G}\bigl(\mathbf{z}(\tau), \tau;\mathbf{E}, \mathbf{B}\bigr) = 0
    \quad
    \Rightarrow
    \quad
    \log f\bigl(\mathbf{z}(t), t\bigr)
    = \log f(\mathbf{z}_0,0).
    \label{eq:vlasov_incompressible}
\end{equation}
Exponentiating both sides yields
\begin{equation}
    f\bigl(\mathbf{z}(t), t\bigr) = f\bigl(\mathbf{z}_0, 0\bigr),
    \label{eq:vlasov_const_along_char}
\end{equation}
revealing that the distribution $f$ is constant along the characteristic flow $\mathbf{z}(t)$. This directly matches the well-known statement of the collisionless Vlasov equation, namely that phase-space density remains constant when particles evolve purely under the Lorentz force.

\subsection{Augmented Flow: Collisional and Additional Physics}
\label{sec:augmented_flow}
While the collisionless Vlasov--Maxwell system (Sec.~\ref{sec:cnf_analogy}) provides a fundamental building block for many plasma simulations, real-world plasmas often exhibit a variety of additional physical processes that alter phase-space dynamics. In this section, we illustrate how such processes can be systematically incorporated into the ODE-based flow perspective. We first discuss collisional effects, which are critical in high-density plasmas (e.g., inertial confinement fusion) \cite{atzeni2004physics} and are frequently handled by Monte Carlo methods outside the main Vlasov solver in typical PIC codes. We then show how other physics, such as radiation reaction, can also be merged into the same framework as long as it can be expressed as an ODE contribution.

\subsubsection{Collisional Effects via Bhatnagar-Gross-Krook (BGK) operator}
The general form of the Fokker-Planck equation for a distribution function $f(\mathbf{x}, \mathbf{v}, t)$ can be written as:
\begin{equation}
\frac{\partial f}{\partial t} + \mathcal{L}{\text{V}} f = \mathcal{C}[f],
\label{eq:fokker_planck}
\end{equation}
where $\mathcal{L}{\text{V}}$ is the Vlasov operator defined in Section~\ref{sec:equivalence_vlasov_particle}, and $\mathcal{C}[f]$ is the collision operator.

The BGK collision operator \cite{bhatnagar1954model} provides a simplified yet physically meaningful approximation to the full collision operator. It models collisions through relaxation to a local equilibrium distribution $f_0$ with a characteristic collision frequency $\nu$:
\begin{equation}
\mathcal{C}_{\text{BGK}}[f] = -\nu(f - f_0),
\label{eq:bgk_operator}
\end{equation}
where $f_0$ is a local Maxwellian distribution:
\begin{equation}
f_0(\mathbf{x},\mathbf{v},t) = N \left(\frac{m}{2\pi k_B T(\mathbf{x},t)}\right)^{3/2}\exp\left(-\frac{m(\mathbf{v}-\mathbf{u}(\mathbf{x},t))^2}{2k_B T(\mathbf{x},t)}\right).
\label{eq:maxwellian}
\end{equation}
The local bulk properties $\mathbf{u}(\mathbf{x},t)$ and $T(\mathbf{x},t)$ are determined from velocity moments of the distribution function:
\begin{align}
\mathbf{u}(\mathbf{x},t) &= \frac{1}{N}\int \mathbf{v}f(\mathbf{x},\mathbf{v},t),d\mathbf{v}, \label{eq:velocity_moment}\\
T(\mathbf{x},t) &= \frac{m}{3k_B N}\int |\mathbf{v}-\mathbf{u}|^2f(\mathbf{x},\mathbf{v},t),d\mathbf{v}. \label{eq:temperature_moment}
\end{align}

In the context of our flow-based framework, where each marker carries its own distribution, the implementation of the BGK operator becomes particularly straightforward. At each timestep, the local equilibrium function $f_0$ can be constructed directly from the particle distributions by computing the necessary moments. Since $f_0$ takes the form of a normalized Gaussian, its value at each marker's phase space position is readily obtained once the bulk velocity and temperature are known. The evolution of the distribution function then follows a simple relaxation equation outside the characteristic flow:
\begin{equation}
\frac{df(z)}{dt} = -\nu(f(z) - f_0(z)),
\label{eq:bgk_evolution}
\end{equation}
where $z$ represents the phase space coordinates. This equation can be solved independently of the flow structure, providing an instantaneous update to the distribution function.

\subsubsection{Extension to Additional Physics}
The flow-based framework naturally extends to incorporate additional physical processes when they can be expressed as ODEs in terms of particle dynamics. Consider a general system where the phase space evolution is governed by both the standard electromagnetic forces and additional physics contributions, collectively denoted by $\mathbf{G}(\mathbf{z},t)$. The complete dynamics can then be described by a coupled system of equations:
\begin{align}
\frac{d\mathbf{z}}{dt} &= \mathbf{G}(\mathbf{z},t), \label{eq:extended_particle_flow} \\
\frac{df(\mathbf{z})}{dt} &= -f(\mathbf{z}) \nabla_{\mathbf{z}}\cdot\mathbf{G}(\mathbf{z},t) - \nu(f(\mathbf{z})-f_0(\mathbf{z})). \label{eq:extended_distribution_flow}
\end{align}
Here, Eq.~\eqref{eq:extended_distribution_flow} captures both the instantaneous change of variables (first term) and the collisional relaxation (second term). This formulation maintains the essential structure of the flow method while accommodating both the phase space compression effects and collisional physics.

The decomposition into marker flow and distribution evolution preserves the computational advantages of the original method while extending its physical scope. Additional effects such as radiation reaction, quantum corrections, or other microscopic processes can be incorporated by appropriate modification of the force term $\mathbf{G}(\mathbf{z},t)$, with their phase space compression effects automatically captured through the divergence term.

\paragraph{Addressing the Reversibility of Collisions}
\quad

A common concern is that collisions, being physically irreversible from a macroscopic perspective, may violate the continuously reversible transformation assumption inherent to our flow-based framework. However, two viewpoints clarify why this does not pose a fundamental problem:

\begin{enumerate}
    \item \textbf{Collisions handled outside the flow structure.}  
    One can treat the collision-induced distribution updates \textit{externally}, i.e., view the continuous flow as evolving in segments. After each collision update, the flow restarts with a new initial condition. This approach is analogous to Particle-In-Cell (PIC) methods, which handle collisions separately from the Newton--Lorentz equations, thereby keeping the particle trajectories continuous \textit{between} collisions.

    \item \textbf{Collisions embedded in a single continuous flow.}  
    In contrast to PIC, it is still possible to incorporate collisions \textit{within} a single continuous flow by formulating an appropriate ODE that approximates collision effects arbitrarily well. At the microscopic (particle) level, collisions are mathematically reversible. Given particle positions and the distribution function, one can recover the initial state by integrating the flow equations (e.g., Eq.~\eqref{eq:extended_particle_flow} and Eq.~\eqref{eq:extended_distribution_flow}) backward in time. Thus, while collisions appear irreversible on a macroscopic scale (due to thermodynamic considerations), they remain fundamentally reversible when viewed microscopically, ensuring that the core flow-based formulation remains valid.
\end{enumerate}

\section{Numerical Method}

\subsection{Initial Setup and Point Selection}

We begin by determining the simulation domain \(\Omega_x \times \Omega_v\) from the physical problem under consideration. We then choose an initial distribution function \(f_0(x,v)\) for each species (e.g., ions and electrons). A typical choice is a combination of uniform and Gaussian distributions. For each species, the initial distribution must be normalized:
\begin{equation}
    f_0(x,v) = N_0 \, f_0^{\mathrm{norm}}(x,v),
\end{equation}
where
\begin{equation}
    \int_{\Omega_x} \int_{\Omega_v} f_0^{\mathrm{norm}}(x,v)\,\mathrm{d}x\,\mathrm{d}v = 1.
\end{equation}

To discretize this distribution, we sample a set of points \(\{(x_i, v_i)\}\subset S_0\) that sufficiently cover the chosen domain. These points need \emph{not} be drawn from the distribution itself; in fact, quasi-random point sets that uniformly cover the computational region often facilitate more accurate numerical integration. For compactly supported distributions, the points span the support boundary; for unbounded (e.g., Gaussian) distributions, we typically truncate at a prescribed multiple of the characteristic width (e.g., \(3\sigma\)–\(4\sigma\)). 

At each point \((x_i,v_i)\), we evaluate either \(f_0^{\mathrm{norm}}(x_i, v_i)\) or its logarithm,
\[
    \log \bigl[f_0^{\mathrm{norm}}(x_i, v_i)\bigr].
\]

We emphasize that these discrete points are \emph{not} PIC macro-particles. Instead, they represent characteristic initial positions in phase space for the Vlasov equation. Physically, they can be interpreted as “markers”—i.e., single-particle tracers that evolve according to the Newton–Lorentz equations.

\subsection{Time Advancement}

To evolve the distribution function, we must advance the markers in phase space and update the associated values of \(f\) (or \(\log f\)) along the characteristics. Many time-advancement schemes can be used; for clarity, we present a leap-frog method \cite{boris1970relativistic}.

Let \(\mathbf{z}_i(t) = \bigl(\mathbf{x}_i(t), \,\mathbf{v}_i(t)\bigr)\) be the phase-space coordinate of the \(i\)-th marker at time \(t\). Along each characteristic,
\begin{equation}
  \frac{\mathrm{d}}{\mathrm{d}t}\bigl[f(\mathbf{z}_i(t), t)\bigr] \;=\; h \bigl(\mathbf{z}_i(t), t\bigr),
\end{equation}
where \(h\) encapsulates the instantaneous change of variable and any collisional terms.  For a collisionless system in an incompressible phase-space flow \(\nabla_{\mathbf{z}}\!\cdot\!\mathbf{G} = 0\), we have \(h = 0\).

\paragraph{Step 1: Compute Fields.}
Using the marker positions and velocities at the current timestep \(n\), we compute the charge density and current density:
\begin{align}
   \rho^{n}(x) \;=&\; \int_{\Omega_v} q\,\bigl[f^{n}_{\mathrm{ion}}(x,v) \;-\; f^{n}_{\mathrm{electron}}(x,v)\bigr] \,\mathrm{d}v, \\
   \mathbf{J}^{n}(x) \;=&\; \int_{\Omega_v} q\,\mathbf{v}\,\bigl[f^{n}_{\mathrm{ion}}(x,v) \;-\; f^{n}_{\mathrm{electron}}(x,v)\bigr] \,\mathrm{d}v.
\end{align}
The local bulk velocity \(\mathbf{u}^n(\mathbf{x})\) and temperature \(T^n(\mathbf{x})\) are computed similarly:
\begin{align}
\mathbf{u}^n(\mathbf{x}) \;=\;& \frac{1}{N_0} \int_{\Omega_v} \mathbf{v}\,f^n(\mathbf{x},\mathbf{v}) \,\mathrm{d}\mathbf{v}, \label{eq:velocity_moment_numerical}\\[6pt]
T^n(\mathbf{x}) \;=\;& \frac{m}{3k_B\,N_0} \int_{\Omega_v} \bigl|\mathbf{v} - \mathbf{u}^n(\mathbf{x})\bigr|^2\,f^n(\mathbf{x},\mathbf{v}) \,\mathrm{d}\mathbf{v}. \label{eq:temperature_moment_numerical}
\end{align}
To carry out these integrals, we employ a scatter-based scheme akin to Smoothed Particle Hydrodynamics (SPH). The electromagnetic fields \(\mathbf{E}^{n}\) and \(\mathbf{B}^{n}\) at timestep \(n\) then follow from Maxwell’s equations with these sources.

\paragraph{Step 2: Velocity Update.}
The velocity at the half-step \((n+1/2)\) is computed from the half-step \((n-1/2)\) via the Lorentz force:
\begin{equation}
   \mathbf{v}_i^{\,n+\tfrac{1}{2}}
   \;=\; \mathbf{v}_i^{\,n-\tfrac{1}{2}} 
         + \Delta t \,\frac{q}{m} \Bigl[\mathbf{E}^n(\mathbf{x}_i^n) \;+\; \mathbf{v}_i^{\,n}\,\times\,\mathbf{B}^n(\mathbf{x}_i^n)\Bigr],
\end{equation}
where \(\mathbf{v}_i^n = \tfrac12\bigl(\mathbf{v}_i^{\,n-\tfrac{1}{2}} + \mathbf{v}_i^{\,n+\tfrac{1}{2}}\bigr)\).

\paragraph{Step 3: Position Update.}
Once \(\mathbf{v}_i^{\,n+\tfrac{1}{2}}\) is known, the position is updated:

\begin{align}
    \mathbf{x}_i^{\,n+1} &= \mathbf{x}_i^{\,n} \;+\; \mathbf{v}_i^{\,n+\tfrac{1}{2}}\,\Delta t, \\[6pt]
    \mathbf{x}_i^{\,n+\tfrac{1}{2}} &= \tfrac12\bigl(\mathbf{x}_i^{\,n} + \mathbf{x}_i^{\,n+1}\bigr).
\end{align}

\paragraph{Step 4: Distribution Update.}
If \(h\neq 0\), we advance \(f(\mathbf{z}_i)\) as
\begin{equation}
    \label{eq:forward_distribution_update}
    f^{\,n+1} \;=\; f^{\,n} \;+\; \Delta t\,h\bigl(\mathbf{x}_i^{\,n+\tfrac{1}{2}},\,\mathbf{v}_i^{\,n+\tfrac{1}{2}},\,t^{n+\tfrac{1}{2}}\bigr).
\end{equation}

Repeated application of Steps 1–4 advances the distribution in time. In the next section, we will discuss the numerical integration scheme and interpolation details that ensure accurate field solves and moment calculations.

\subsection{Scatter Point Integration}
\label{sec:scatter_point_integration}

In our method, each marker carries its position \(\mathbf{x}_i\), velocity \(\mathbf{v}_i\), and a value \(f_i\) that represents the phase-space distribution at \(\bigl(\mathbf{x}_i, \mathbf{v}_i\bigr)\). This strategy enables accurate numerical integration rather than standard Monte Carlo (MC) sampling. Crucially, however, the marker locations (and their associated values) cannot be freely chosen at each time step because \(\mathbf{x}_i\) and \(\mathbf{v}_i\) are determined by integrating the governing ODEs forward in time from previously known states.

If one wishes to add markers at arbitrary positions, it is necessary to integrate backward in time from the desired terminal position. In principle, backward-in-time integration is less computationally demanding than forward integration because it can reuse previously computed fields \(\mathbf{E}\) and \(\mathbf{B}\). Only the saved field data from past time steps are needed to back-trace trajectories and update the corresponding values \(f_i\). In practice, however, performing this procedure every time step is still too costly. As a result, we typically rely on the existing, forward-propagated markers to advance the solution. Nevertheless, the ability to spawn new markers on demand can be leveraged for adaptive refinement, as discussed in the subsection \ref{sec:adaptive_refinement}.

Because markers cannot be freely relocated, standard Gaussian quadrature or uniform grid-based integration \cite{davis2007methods} is not directly applicable. Following the rationale of smoothed particle hydrodynamics (SPH), we assign each marker a compactly supported kernel \(\phi\) with characteristic size \(h\). Common choices include a cubic spline or other radially symmetric kernels \cite{monaghan1992smoothed}. In practice, for optimal results, one must nondimensionalize carefully and adapt \(h\) based on local marker distribution. One option is
\[
  h_i \;=\; k\,d_{\min}\bigl(\mathbf{x}_i, \mathbf{v}_i\bigr),
\]
where \(d_{\min}(\mathbf{x}_i, \mathbf{v}_i)\) is the distance in phase space to the nearest neighbor, and \(k>1\) is a user-defined constant.

\subsubsection{Forming the phase-space interpolation}

Given \(N\) markers, we define a continuous approximation to the phase-space distribution via the kernel-weighted interpolation. $\phi\bigl(\mathbf{x},\mathbf{v}\mid \mathbf{x}_i,\mathbf{v}_i\bigr)$ refers to the kernel function centered at $\mathbf{x}_i,\mathbf{v}_i$.
\begin{equation}
  f(\mathbf{x},\mathbf{v}) 
  \;=\;
  \frac{\displaystyle \sum_{i=1}^{N} f_i \,\phi\bigl(\mathbf{x},\mathbf{v}\mid \mathbf{x}_i,\mathbf{v}_i\bigr)}
       {\displaystyle \sum_{i=1}^{N} \phi\bigl(\mathbf{x},\mathbf{v}\mid \mathbf{x}_i,\mathbf{v}_i\bigr)},
  \label{eq:phase_space_interpolation}
\end{equation}
which then allows the recovery of scalar or vector moments of \(f\). For a plasma consisting of ions (ion) and electrons (elec), the charge density \(\rho(\mathbf{x})\) and current density \(\mathbf{J}(\mathbf{x})\) can be written as
\begin{subequations}
\label{eq:rho_J_defs}
\begin{align}
  \rho(\mathbf{x}) 
  &= q
  \int \!\Biggl[ 
   \frac{\displaystyle \sum_{i=1}^{N} f_i^{(\mathrm{ion})} \,\phi\bigl(\mathbf{x},\mathbf{v}\mid \mathbf{x}_i,\mathbf{v}_i\bigr)}
        {\displaystyle \sum_{i=1}^{N} \phi\bigl(\mathbf{x},\mathbf{v}\mid \mathbf{x}_i,\mathbf{v}_i\bigr)}
   \;-\;
   \frac{\displaystyle \sum_{i=1}^{N} f_i^{(\mathrm{elec})} \,\phi\bigl(\mathbf{x},\mathbf{v}\mid \mathbf{x}_i,\mathbf{v}_i\bigr)}
        {\displaystyle \sum_{i=1}^{N} \phi\bigl(\mathbf{x},\mathbf{v}\mid \mathbf{x}_i,\mathbf{v}_i\bigr)}
  \Biggr] \mathrm{d}\mathbf{v}, \\[6pt]
  \mathbf{J}(\mathbf{x}) 
  &= q
  \int \!\Biggl[ 
   \frac{\displaystyle \sum_{i=1}^{N} \mathbf{v}_i\, f_i^{(\mathrm{ion})} \,\phi\bigl(\mathbf{x},\mathbf{v}\mid \mathbf{x}_i,\mathbf{v}_i\bigr)}
        {\displaystyle \sum_{i=1}^{N} \phi\bigl(\mathbf{x},\mathbf{v}\mid \mathbf{x}_i,\mathbf{v}_i\bigr)}
   \;-\;
   \frac{\displaystyle \sum_{i=1}^{N} \mathbf{v}_i\, f_i^{(\mathrm{elec})} \,\phi\bigl(\mathbf{x},\mathbf{v}\mid \mathbf{x}_i,\mathbf{v}_i\bigr)}
        {\displaystyle \sum_{i=1}^{N} \phi\bigl(\mathbf{x},\mathbf{v}\mid \mathbf{x}_i,\mathbf{v}_i\bigr)}
  \Biggr] \mathrm{d}\mathbf{v}.
\end{align}
\end{subequations}

\subsubsection{Quasi-static Field Solution via Green Method}

Instead of reconstructing $f$ on a velocity mesh, we can compute $\mathbf{E}$ and $\mathbf{B}$ directly through integral equations. For many plasma applications (e.g., tokamaks, Z-pinches), the conduction current dominates the displacement current, allowing us to employ the \emph{quasi-static} approximation\cite{freidberg2014ideal}. This approximation holds when $\epsilon_0 \,\partial \mathbf{E}/\partial t \ll \mathbf{J}$, making the displacement-current term in the Ampère--Maxwell law negligible. Under these conditions, Maxwell's equations reduce to:
\begin{equation}
  \nabla \cdot \mathbf{E} = \frac{\rho}{\epsilon_0},
  \quad
  \nabla \times \mathbf{B} = \mu_0 \,\mathbf{J}.
  \label{eq:maxwells_quasistatic}
\end{equation}
Using Green's functions, the solutions of Electric and magnetic potential take the form:
\begin{equation}
  \mathbf{\phi}(\mathbf{x})
  =
  \frac{1}{4 \pi \epsilon_0}
  \int
  \rho(\mathbf{x}')\,G_E(\mathbf{x},\mathbf{x}')
  \,\mathrm{d}V'+
  \frac{1}{4 \pi} \oint
  \biggl[
  G_E(\mathbf{x},\mathbf{x}') \frac{\partial \phi}{\partial n'}-
  \phi \frac{\partial G_E(\mathbf{x},\mathbf{x}')}{\partial n'} 
  \biggr]
  dS'
\end{equation}

\begin{equation}
\begin{split}
\mathbf{A}(\mathbf{x})
&=
\frac{\mu_0}{4\pi}
\int_V 
\mathbf{J}(\mathbf{x}')\,G_B(\mathbf{x},\mathbf{x}')
\,\mathrm{d}V' \\
&\quad+\;
\frac{1}{4\pi}
\oint_S
\biggl[
G_B(\mathbf{x},\mathbf{x}')\,\frac{\partial \mathbf{A}}{\partial n'}(\mathbf{x}')
\;-\;
\mathbf{A}(\mathbf{x}')\,\frac{\partial G_B(\mathbf{x},\mathbf{x}')}{\partial n'}
\biggr]
\,\mathrm{d}S'.
\end{split}
\end{equation}

where $G_E$ and $G_B$ are Green's functions satisfying the appropriate boundary conditions. Applying Green's theorem \cite{jackson2021classical} separates these solutions into volume and boundary (surface) terms. In plasma physics simulations, the boundary terms can often be computed independently of plasma dynamics and imposed on the domain through separate methods.

For illustration, we present the free-space case where boundary terms vanish. The general case follows the same procedure but requires Green's functions chosen to match the specific boundary conditions and the geometry of the computational domain:
\begin{subequations}
\label{eq:potential_to_field}
\begin{align}
\nabla \cdot \phi &= \mathbf{E}, \\
\nabla \times \mathbf{A} &= \mathbf{B}.
\end{align}
\end{subequations}

\begin{subequations}
\label{eq:E_B_integral_solutions_free}
\begin{align}
  \mathbf{E}(\mathbf{x})
  &= 
  \frac{1}{4\pi \,\epsilon_0}
  \int 
    \frac{\rho(\mathbf{x}')\,\bigl[\mathbf{x}-\mathbf{x}'\bigr]}
         {\|\mathbf{x}-\mathbf{x}'\|^3}
  \,\mathrm{d}V',
  \\[6pt]
  \mathbf{B}(\mathbf{x})
  &= 
  \frac{\mu_0}{4\pi}
  \int 
    \frac{\mathbf{J}(\mathbf{x}') \times \bigl[\mathbf{x}-\mathbf{x}'\bigr]}
         {\|\mathbf{x}-\mathbf{x}'\|^3}
  \,\mathrm{d}V' .
\end{align}
\end{subequations}
Substituting $\rho$ and $\mathbf{J}$ from Eq.~\eqref{eq:rho_J_defs} yields integrals over both configuration space ($\mathrm{d}V'$) and velocity space ($\mathrm{d}\mathbf{v}$). The compact support of each kernel $\phi$ localizes these integrals, reducing computational complexity.

\paragraph{Evaluating the kernel integrals.}\quad

We wish to compute the electric field \(\mathbf{E}(\mathbf{x}_i)\) at a marker’s location \(\mathbf{x}_i\). A typical scalar component of this field has the form
\begin{equation}
  E(\mathbf{x}_i)
  = \sum_{k=1}^{N} f_k 
  \int
    \frac{\phi\bigl(\mathbf{x}',\mathbf{v}' \mid \mathbf{x}_k,\mathbf{v}_k\bigr)}
         {\|\mathbf{x}'-\mathbf{x}_i\|^2 
          \,\Bigl[\sum_{j=1}^{N} \phi\bigl(\mathbf{x}',\mathbf{v}' \mid \mathbf{x}_j,\mathbf{v}_j\bigr)\Bigr]}
  \,\mathrm{d}V'\,\mathrm{d}\mathbf{v}' 
  + \dots
\end{equation}
Because the kernel \(\phi\) is compactly supported, each volume integral is confined to a (finite) domain around \(\mathbf{x}_k\). Hence, only \(N\) such integrals appear in the summation.

A concern arises when \(\mathbf{x}_i\) lies inside—or very near—one of the kernel supports. The factor \(\|\mathbf{x}' - \mathbf{x}_i\|^{-2}\) can appear “singular” near \(\mathbf{x}' = \mathbf{x}_i\). Nevertheless, it can be shown that this integral remains finite and, crucially, can be approximated well by standard Gaussian quadrature. Below, we sketch the core idea of the proof.

    Suppose \(\mathbf{x}_i\) lies within (or on the boundary of) the support of \(\phi\). At worst, there is at most one integrand term of the form
    \[
      \frac{1}{\|\mathbf{x}' - \mathbf{x}_i\|^2}
    \]
    that becomes large if \(\mathbf{x}' \approx \mathbf{x}_i\).

    Let \(B_\epsilon(\mathbf{x}_i)\) be a small ball (sphere) of radius \(\epsilon\) centered at \(\mathbf{x}_i\). We split the integral over the kernel’s support into two parts:

\begin{itemize}
    \item \textbf{Outside} \(B_\epsilon(\mathbf{x}_i)\), the integrand is smooth (no singularity), so standard quadrature applies directly.
    \item \textbf{Inside} \(B_\epsilon(\mathbf{x}_i)\), we examine the behavior of
      \(\|\mathbf{x}' - \mathbf{x}_i\|^{-2}\).

\end{itemize}

Inside \(B_\epsilon(\mathbf{x}_i)\), observed that the rest of the integrand (including \(\phi\) and the denominator’s sum over \(\phi\)) is approximately constant over that ball when $\epsilon$ is taken sufficiently small, taking some representative value \(C\). The problematic term simplifies to: (neglecting velocity integration)
    \[
      \int_{\|\mathbf{x}' - \mathbf{x}_i\|\le \epsilon} \frac{C}{\|\mathbf{x}'-\mathbf{x}_i\|^2}\,\mathrm{d}V'
      = C \int_0^\epsilon \!\!\int_{\phi}\int_{\theta} \frac{1}{r^2} r^2 sin(\theta)\,\mathrm{d}\phi d\theta \,\mathrm{d}r,
    \]

    which \textbf{goes to zero} linearly with \(\epsilon\). Therefore,
    \[
      \int_{\|\mathbf{x}' - \mathbf{x}_i\|\le \epsilon} \frac{C}{\|\mathbf{x}'-\mathbf{x}_i\|^2}\,\mathrm{d}V'
      \approx 4\pi\,C\,\epsilon \to 0 
      \quad\text{as}\quad \epsilon \to 0.
    \]

    Because the integral is finite (in fact, tends to zero for sufficiently small \(\epsilon\)), a standard Gaussian quadrature rule on a well-resolved mesh inside the kernel support will yield a convergent approximation. In practice, we simply ensure that no quadrature point is placed \textit{exactly} at the singularity; the \(1/\|\mathbf{x}'-\mathbf{x}_i\|^2\) factor is still well-behaved in an integral sense.

\paragraph{Complexity Reduction in Field Computation}\quad

Suppose we use \(N_{\mathrm{quad}}\) quadrature points per kernel support. Naïvely, evaluating the field at \(N\) marker positions suggests an \(O\bigl(N^2\,N_{\mathrm{quad}}\bigr)\) cost. However, a key optimization dramatically reduces this cost.

Observe that in field computations the integrand can be separated into a \emph{distribution-weighted} part and a \emph{geometric} factor:
\begin{equation}
  \int 
    \underbrace{\biggl[
    \frac{\phi\bigl(\mathbf{x}',\mathbf{v}' \mid \mathbf{x}_k,\mathbf{v}_k\bigr)}
         {\sum_{j=1}^{N} \phi\bigl(\mathbf{x}',\mathbf{v}' \mid \mathbf{x}_j,\mathbf{v}_j\bigr)}
    \biggr]}_{\text{distribution-weighted factor}}
    \;\times\;
    \underbrace{G\!\Bigl(\|\mathbf{x}-\mathbf{x}'\|\Bigr)}_{\text{geometric factor}}
  \,\mathrm{d}V'\,\mathrm{d}\mathbf{v}',
  \label{eq:common_part_geometric_part}
\end{equation}
where the bracketed term depends only on how markers’ kernels combine locally, while \(G(\|\mathbf{x}-\mathbf{x}'\|)\) (e.g.,  \([\mathbf{x}-\mathbf{x}']/\|\mathbf{x}-\mathbf{x}'\|^3\)) depends on the chosen evaluation point and the integration variable.

Because the distribution-weighted part depends only on local kernel contributions, it can be computed once and reused. Forming this \emph{common} distribution-weighted data for \(N\) markers and \(N_{\mathrm{quad}}\) quadrature nodes in each support costs \(O\bigl(N\,N_{\mathrm{quad}}\bigr)\). Specifically, one tabulates
\[
  \frac{\phi(\mathbf{x}',\mathbf{v}' \mid \mathbf{x}_k,\mathbf{v}_k)}
         {\sum_{j=1}^{N} \phi(\mathbf{x}',\mathbf{v}' \mid \mathbf{x}_j,\mathbf{v}_j)}
\]
at the \(N_{\mathrm{quad}}\) nodes within each of the \(N\) marker supports. Once computed, these tabulated values can be multiplied by any geometric factor \(G(\|\mathbf{x}-\mathbf{x}'\|)\) to evaluate \(\mathbf{E}\) or \(\mathbf{B}\) at a new point \(\mathbf{x}\) by a simple sum over the quadrature nodes. Hence, after this one-time \(O\bigl(N\,N_{\mathrm{quad}}\bigr)\) construction, field evaluation at each new \(\mathbf{x}_i\) becomes much cheaper.

In practice, the field often needs to be evaluated only on a grid of \(M\) points. Due to the rapid decay of \(G(\|\mathbf{x}-\mathbf{x}'\|)\), each evaluation point effectively interacts with only \(N_{\mathrm{neighbor}}\) significant marker supports. Thus, without separating out the reusable distribution-weighted part, the worst-case cost to evaluate the field at \(M\) points is \(O(M \, N_{\mathrm{quad}} \, N_{\mathrm{neighbor}})\). 

\emph{Which approach is preferable depends on the scenario}: if one needs to evaluate the field many times after forming the distribution, precomputing the common part offers large savings. If only a few evaluations are required, computing distribution and geometry on the fly may be cheaper.

\subsubsection{Full Field Solution}

In the full-field approach, unlike the quasi-static case, it is not possible to directly obtain the field solution from an integral using Green’s functions. Instead, one must first compute \(\rho\) and \(\mathbf{J}\) from the distribution and then solve for the fields using a standard field solver, similar to the approach used in Particle-In-Cell (PIC) methods.

To evaluate \(\rho\) and \(\mathbf{J}\) at each spatial grid point, we employ Gaussian quadrature in velocity space. Specifically, for each physical grid point, \(N_{\text{quad}}\) Gaussian points in velocity space are used to perform the marginal integration. The distribution function at each of these Gaussian points is obtained by interpolation via Eq.~\eqref{eq:phase_space_interpolation}.

Once \(\rho\) and \(\mathbf{J}\) have been calculated on the grid, the fields can be solved using any standard field solver. We refer to this as the “full field” solution because it captures all field contributions beyond the quasi-static approximation. However, if desired, this approach could recover the quasi-static solution by employing a quasi-static solver. We do not provide details of the specific field solver here.

The dominant computational cost of this method arises from the calculation of \(\rho\) and \(\mathbf{J}\), which scales as \(O(M \times N_{\text{quad}})\), where \(M\) is the number of spatial grid points and \(N_{\text{quad}}\) is the number of quadrature points used in velocity space.

This full-field approach is particularly beneficial in scenarios where the quasi-static approximation fails or where boundary conditions and computational domains are too complex for Green’s function-based methods. Furthermore, in cases with relatively low dimensionality in velocity space, the full-field solution can be more efficient than a Green’s function-based quasi-static approach.

\subsubsection{Error Estimation}
\label{sec:Error Estimation}
An analytical evaluation of the method's error would be challenging, if not impossible, as it depends on various functions and point placements. However, we can analyze a special case: the standard normalization in $N_d$-dimensional space using our method with $N_p$ points, evaluating the interpolation error on a regular grid.
Through empirical testing across different dimensions and point counts, we find that the error follows the relation $c/N_p^m$, where:
\begin{equation}
m \approx \frac{2}{N_d+1}
\end{equation}
Given that the average distance between points is:
\begin{equation}
h \propto (1/N_p)^{1/N_d}
\end{equation}
The \emph{interpolation} error is therefore:
\begin{equation}
O(h^{\frac{2N_d}{N_d+1}})
\end{equation}
Since the integration method is of higher order than interpolation, and the deduction of the value involves full integration, the integration error becomes:
\begin{equation}
O(h^{\frac{2N_d}{N_d+1}+N_d}) = O(h^{\frac{3N_d+N_d}{N_d+1}})
\end{equation}
Converting back to $N_p$:
\begin{equation}
O(\frac{1}{N_p}^{\frac{N_d+3}{N_d+1}}) = O(\frac{1}{N_p}^{1+\frac{2}{N_d+1}})
\end{equation}
While the method becomes less efficient as dimensionality increases, the convergence rate remains greater than 1. It should be noted that this represents an idealized case; in practice, with more complex distributions, the convergence rate may be slower.

\subsubsection{Adaptive Refinement via Marker Spawning and Backward Integration}
\label{sec:adaptive_refinement}

Even though the scatter-point method primarily employs forward-integrated markers to resolve the phase-space distribution, certain regions may develop steep gradients or localized structures that remain under-resolved. In such scenarios, \emph{adaptive refinement} through the spawning of new markers can substantially enhance numerical fidelity.

\paragraph{Spawning new markers.}
The key idea behind marker spawning is to place a new marker at a desired position--velocity pair, 
\(\bigl(\mathbf{x}_{\star}, \mathbf{v}_{\star}\bigr)\), precisely where higher resolution is needed. Because our governing equations are characteristic-based, the new marker’s position and velocity cannot be assigned arbitrarily at the final time; instead, we \emph{back-trace} its trajectory to an earlier time to ensure consistency with the solution’s advective nature:

\begin{enumerate}
  \item \textbf{Backward Integration (to \(t=0\)):}  
  Integrate the characteristic equations in reverse from 
  \(\bigl(\mathbf{x}_{\star}, \mathbf{v}_{\star}\bigr)\) (current time \(t\)) back to 
  \(\bigl(\mathbf{x}_{0}, \mathbf{v}_{0}\bigr)\) at \(t=0\). This step is computationally inexpensive because the required fields are already recorded and do not need to be recalculated.

  \item \textbf{Initial Distribution Lookup:}  
  Since \(f\) is fully specified at \(t=0\), retrieve  
  \[
    f_{0} \;=\; f\bigl(\mathbf{x}_{0}, \mathbf{v}_{0}, 0\bigr).
  \]

  \item \textbf{Forward Continuous Transformation (to obtain \(f_{\star}\)):}  
  Use a continuous change-of-variables method to evolve \(f_{0}\) from 
  \(\bigl(\mathbf{x}_{0}, \mathbf{v}_{0}\bigr)\) forward to 
  \(\bigl(\mathbf{x}_{\star}, \mathbf{v}_{\star}\bigr)\). The final value, \(f_{\star}\), follows from the instantaneous change-of-variables formula and any collisional terms (as in Eq.~\eqref{eq:forward_distribution_update}). 
\end{enumerate}

Once the backward and forward integrations are completed, the \emph{newly spawned} marker inherits its updated value \(f_{\star}\) and is inserted into the forward integration to continue the solution alongside the rest of the marker ensemble.

\paragraph{Refinement criteria.}
To keep the additional overhead manageable, new markers should only be spawned where they are most needed. Possible refinement indicators include:

\textbf{Gradient-based triggers.}  
  Identify regions of large \(\nabla f\) and spawn additional markers to capture steep gradients.

\textbf{Kernel support checks.}  
  If $h_i$ for certain kernel becomes too large, indicating insufficient local resolution, additional markers may be inserted.

\paragraph{Cost considerations and kernel adaptation.}
Though backward integration for a small set of newly spawned markers is economical, it becomes expensive if performed at every time step. In practice, one refines adaptively and \emph{infrequently}, only in response to the above triggers. Newly created markers are then propagated forward as part of the ensemble without further backward steps. When markers are introduced in a dense region of phase space, their kernel size \(h_i\) can be adapted accordingly.

\subsection{Marker Kernel Distance and Time Step Criteria}
\label{sec:marker-kernel-distance-time-step}

When using spherical or hyperspherical kernels to represent marker distributions in phase space, it is crucial to ensure that the spatial and velocity dimensions share comparable characteristic lengths. Often, one non-dimensionalizes \cite{verboncoeur2005particle} the problem so that the kernel has a uniform radius in both physical and velocity spaces. However, if significantly different characteristic scales exist (e.g., due to anisotropic temperature or disparate spatial domains), one can instead adopt a \emph{non-uniform} kernel that varies spatially or with velocity to accommodate these variations.

\paragraph{Kernel Size Bounds.}
We consider two principal bounds on the kernel size (or \emph{radius}) in physical space:
\begin{enumerate}
  \item \textbf{Debye Length Constraint.}  
    In many kinetic plasma simulations, the maximum kernel radius in physical space is naturally capped by the Debye length \(\lambda_D\). Since \(\lambda_D\) characterizes the scale over which electric potential perturbations are screened out, choosing a kernel radius much larger than \(\lambda_D\) risks smoothing out physically important electrostatic effects.

  \item \textbf{Velocity Distribution Resolution.}  
    A second consideration is the ability of the kernel to capture the local velocity distribution. Typically, a kernel radius \(\delta_v\) in velocity space should be chosen smaller than the local thermal velocity scale \(v_{\mathrm{th}}\). However, because thermal spreads are often less restrictive, this velocity-based bound on the kernel can be more relaxed than the Debye-length constraint.
\end{enumerate}   

\paragraph{Time Step Constraints.}
As in standard continuum Vlasov or particle-in-cell (PIC) simulations, time step \(\Delta t\) of flow-based method is limited by the electron plasma frequency,
\begin{equation}
  \omega_{pe} \;=\; \sqrt{\frac{n_e\,e^2}{\varepsilon_0\,m_e}},
\end{equation}
where \(m_e\) is the electron mass. To resolve plasma oscillations, one typically requires
\begin{equation}
  \Delta t \;\lesssim\; \frac{C}{\omega_{pe}},
  \label{eq:plasma-time-step}
\end{equation}
for some constant \(C \lesssim 1\). This constraint can be more stringent than the usual Courant–Friedrichs–Lewy (CFL) condition that appears in fluid or other wave-like PDEs, especially if the thermal velocity is high.

Overall, the choice of numerical parameters (h and $\Delta t$) in flow-based methods shares similarities with direct Vlasov solvers \cite{cheng1976integration} in managing dispersion and aliasing effects. However, flow-based methods using distribution function markers offer greater flexibility due to their Lagrangian nature: Markers following phase space characteristics naturally adapt their density to the distribution function, providing enhanced resolution in regions of high concentration. This self-adapting behavior allows some relaxation of the strict constraints ($\delta_r < \lambda_D$ and $\Delta t < 1/\omega_{pe}$) typically required in direct Vlasov or PIC simulations. Nevertheless, kernel sizes must still be carefully chosen: small enough to resolve fine-scale phenomena like Debye screening, yet large enough to maintain computational efficiency with a reasonable number of markers.

In summary, for kernel-based marker methods, one typically sets:
\[
 h \;\sim\; \mathcal{O}(\lambda_D), 
  \quad
  \Delta t \;\lesssim\; \frac{C}{\omega_{pe}},
\]

\section{Results: Validation Against Classical Benchmarks}

To validate our approach, we examine three classical plasma physics problems: Landau damping and two-stream instability in collisionless plasma, along with a variation of the two-stream problem for collisional plasma.

All simulations use a fully non-dimensionalized system, where we simulate electron behavior with stationary ion background. In this system, the electron mass and charge are normalized to unity, and the plasma frequency is set to 1, yielding a charge-to-mass ratio $q_m = -1$. For validation purposes, all simulations are performed in a 1D1V configuration to facilitate visualization and comparison with theoretical solutions. 

We compare our results against a baseline Particle-In-Cell (PIC) method, with parameters drawn from standard plasma physics simulations. We adopt the PIC code implementation and plasma parameters from \cite{xie2018computational,lapenta2006particle}. Unless otherwise specified, we employ the leapfrog method for time advancement. In all the demonstrations below, we use a quasi-static field solver. The full field solver yields very similar results to the quasi-static field solver in our testing.
\subsection{Landau Damping}
For the Landau damping case, we set the electron thermal velocity $v_t = 1$. To ensure single-mode analysis, the spatial domain length is set to $L = 2\pi/k$, where $k$ is the perturbation wavenumber. The neutral charge density, derived from the plasma frequency, is:

\begin{equation}
\rho_{elec} = \omega_p^2/q_m = -1
\end{equation}
The total number of markers (electrons or ions) is:
\begin{equation}
N_{tot} = L\omega_p^2/m_{elec} = L
\end{equation}
We initialize the system with a density perturbation resulting in the distribution:

\begin{subequations}
\begin{align}
    f = \frac{N_{tot}(1 + \epsilon \sin(kx))}{L} \frac{1}{\sqrt{2\pi v_t}} \exp\left(-\frac{v^2}{2v_t^2}\right), \\
   f_{norm} = \frac{1 + \epsilon \sin(kx)}{L} \frac{1}{\sqrt{2\pi v_t}} \exp\left(-\frac{v^2}{2v_t^2}\right).
\end{align}
\end{subequations}

where $\epsilon$ represents the perturbation amplitude.

Markers are distributed uniformly in the $x$-$v_x$ plane, with $x \in [0,L]$ and $v_x \in [-3.5v_t, 3.5v_t]$. The velocity bounds are chosen to ensure the integrated distribution approaches unity. For this collisionless case, the distribution value associated with each marker remains constant throughout the simulation.

For comparison, we use 3,500 markers in our method versus 350,000 particles in the PIC simulation. The parameters include: time step $\Delta t = 0.01$, 32 grid points for electric field evaluation, initial perturbation $\epsilon = 0.2$, and perturbation wavenumber $k = 0.7$. We assess the damping by measuring the electric field energy $E_E = \frac{1}{2}\sum E_x^2 \Delta x$ at each time step. By plotting $\log(\sqrt{E_E})$, we can extract both the oscillation frequency $\omega$ and damping rate $\gamma$.

\begin{figure}[H]
\centering
\includegraphics[width=0.8\textwidth]{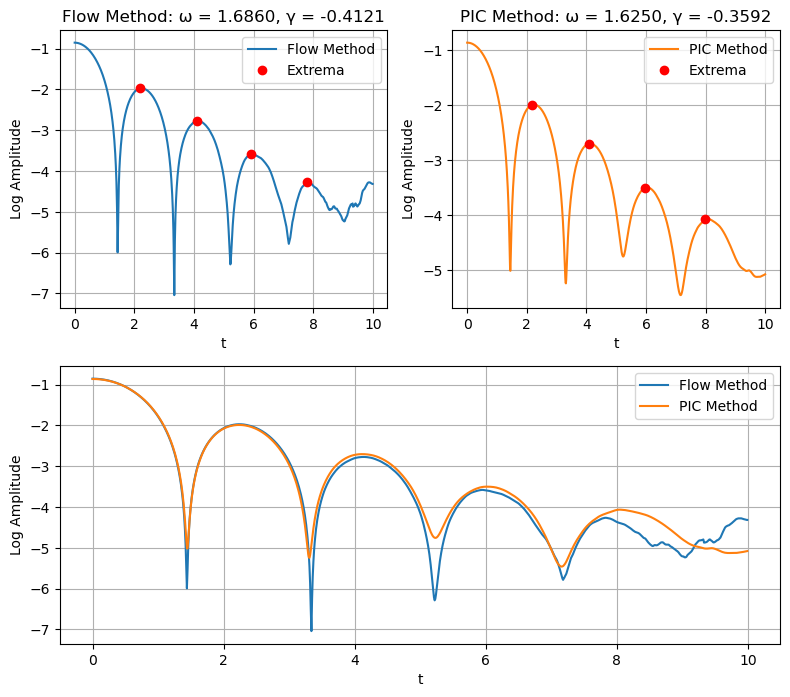}
\caption{Time evolution of electric field energy in Landau damping. Comparison between our flow method and PIC simulation shows comparable accuracy with 100× fewer particles.(theoretical values: $\omega = 1.67387$, $\gamma = -0.392401$)}
\label{Energy_landau}
\end{figure}

The electric field energy evolution is shown in Figure \ref{Energy_landau}. The flow method demonstrates comparable or better accuracy compared to PIC, with both methods reaching similar noise background levels. Our flow method yields $\omega = 1.6860$, $\gamma = -0.4121$, while the PIC method gives $\omega = 1.6250$, $\gamma = -0.3592$ (theoretical values: $\omega = 1.67387$, $\gamma = -0.392401$). Notably, the PIC results show greater sensitivity to background noise.

\begin{figure}[H]
\centering
\includegraphics[width=0.8\textwidth]{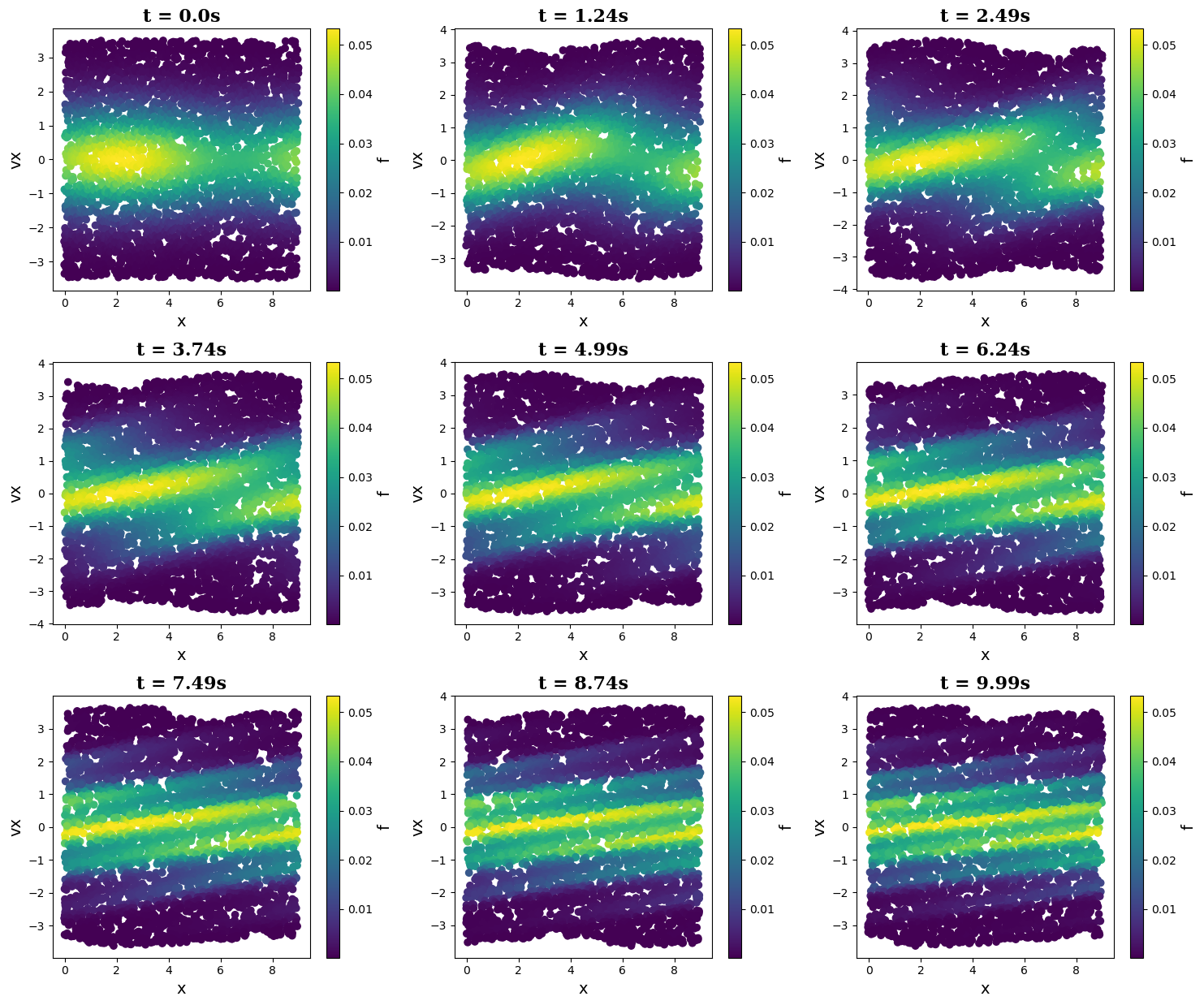}
\caption{Phase space evolution of particle distribution in Landau damping. Colors represent the distribution value of each marker, revealing the detailed phase space structure of the Landau damping process.}
\label{distribution_plot_landau}
\end{figure}

To visualize the phase space dynamics, we plot the marker positions in phase space at different time steps in Figure \ref{distribution_plot_landau}, with colors indicating the distribution value at each point. These phase space plots clearly demonstrate how our method successfully combines aspects of direct Vlasov and particle methods, maintaining accurate distribution information throughout the simulation.

\subsection{Two-Stream Instability}

For the two-stream instability, we simulate two electron beams with bulk velocities $v_b = \pm 1$ and thermal velocity $v_t = 0.3$. The initial distribution includes a velocity perturbation:

\begin{subequations}
\begin{align}
v_{b1} &= v_b + \epsilon \sin(kx) \\
v_{b2} &= -v_b - \epsilon \sin(kx) \\
f_{norm}(x,v) &= \frac{1}{2L}\frac{1}{\sqrt{2\pi v_t}}\left[\exp\left(-\frac{(v-v_{b1})^2}{2v_t^2}\right) + \exp\left(-\frac{(v-v_{b2})^2}{2v_t^2}\right)\right]
\end{align}
\end{subequations}

We compare simulations using 4,000 markers in our method against 400,000 particles in the PIC method, with perturbation amplitude $\epsilon = 0.1$.

\begin{figure}[H]
\centering
\includegraphics[width=0.8\textwidth]{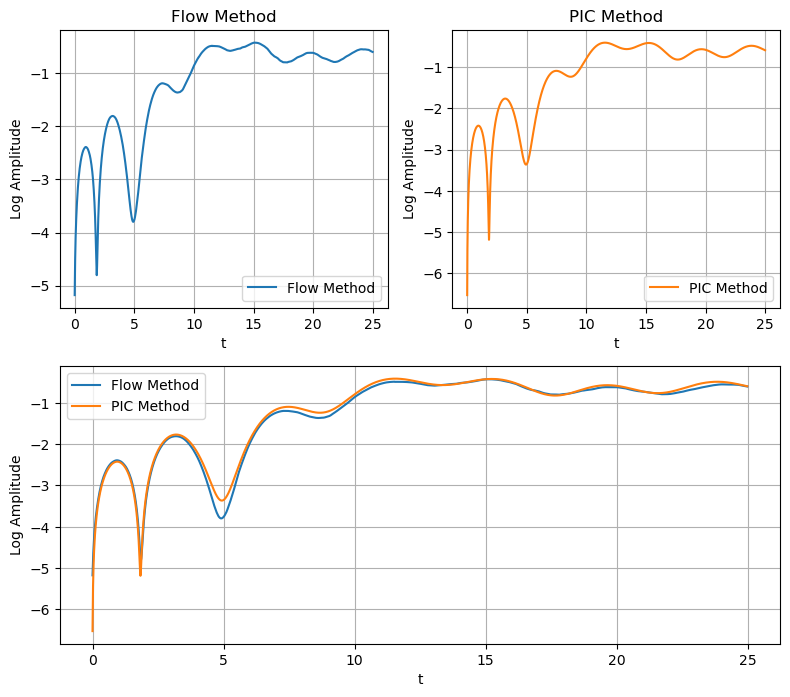}
\caption{Electric field energy evolution in two-stream instability, comparing our method with PIC simulation through both linear growth and saturation phases.}
\label{Energy_2flow}
\end{figure}

The electric field energy evolution for the two-stream instability is presented in Figure \ref{Energy_2flow}. Both methods capture the characteristic stages of the instability, from initial linear growth to nonlinear saturation. The agreement between the methods, despite our approach using 100 times fewer particles, demonstrates the efficiency and accuracy of our method.

\begin{figure}[H]
\centering
\includegraphics[width=0.8\textwidth]{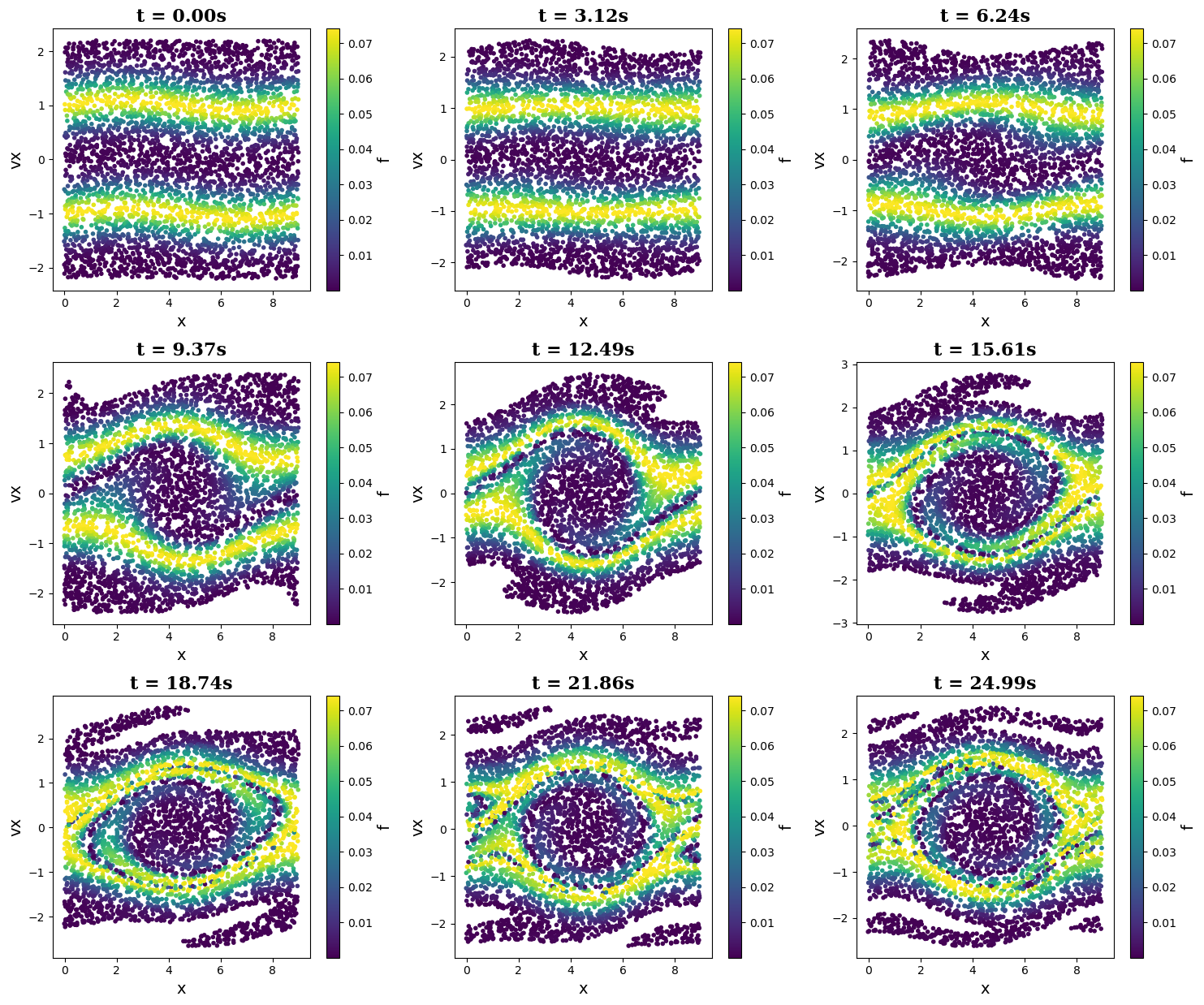}
\caption{Phase space evolution of two-stream instability, showing formation and development of characteristic vortex structures.}
\label{distribution_2flow}
\end{figure}

Figure \ref{distribution_2flow} shows the phase space evolution at different time steps, revealing the formation of the characteristic eye diagrams. Visual inspection of these phase space structures confirms the physical correctness of our results, showing proper development of the instability-driven vortices in phase space.

\subsection{Collisional Relaxation of Two-Stream Distribution}

To validate the collision operator implementation, we examine the relaxation of a two-stream distribution to equilibrium. The initial condition consists of two counter-streaming electron beams without spatial perturbation, which should relax through collisions into a single Maxwellian distribution.

We initialize the system with two symmetric beams having thermal velocity $v_t = 0.3$ and bulk velocities $v_b = \pm 1$. The initial distribution is given by:
\begin{equation}
f_{\text{norm}}(x,v) = \frac{1}{2L}\sum_{i=1}^2\frac{1}{\sqrt{2\pi v_t}}\exp\left(-\frac{(v-(-1)^iv_b)^2}{2v_t^2}\right).
\label{eq:collision_initial}
\end{equation}

The total thermal energy of this system, characterized by the velocity variance $\sigma^2$, should remain constant throughout the collision process due to energy conservation:
\begin{equation}
\sigma^2 = v_b^2 + v_t^2 = 1.09,
\label{eq:energy_conservation}
\end{equation}
corresponding to $\sigma = 1.044$. This value provides a key benchmark for validating the energy conservation properties of our collision operator.

The simulation employs 2,500 markers with a time step $\Delta t = 0.005$ and runs for 3,000 steps. At each step, we compute the local equilibrium distribution $f_0$ using the first two velocity moments and evolve the marker distributions according to Eq.~\eqref{eq:bgk_evolution}.

\begin{figure}[H]
\centering
\begin{subfigure}{0.48\textwidth}
\includegraphics[width=\textwidth]{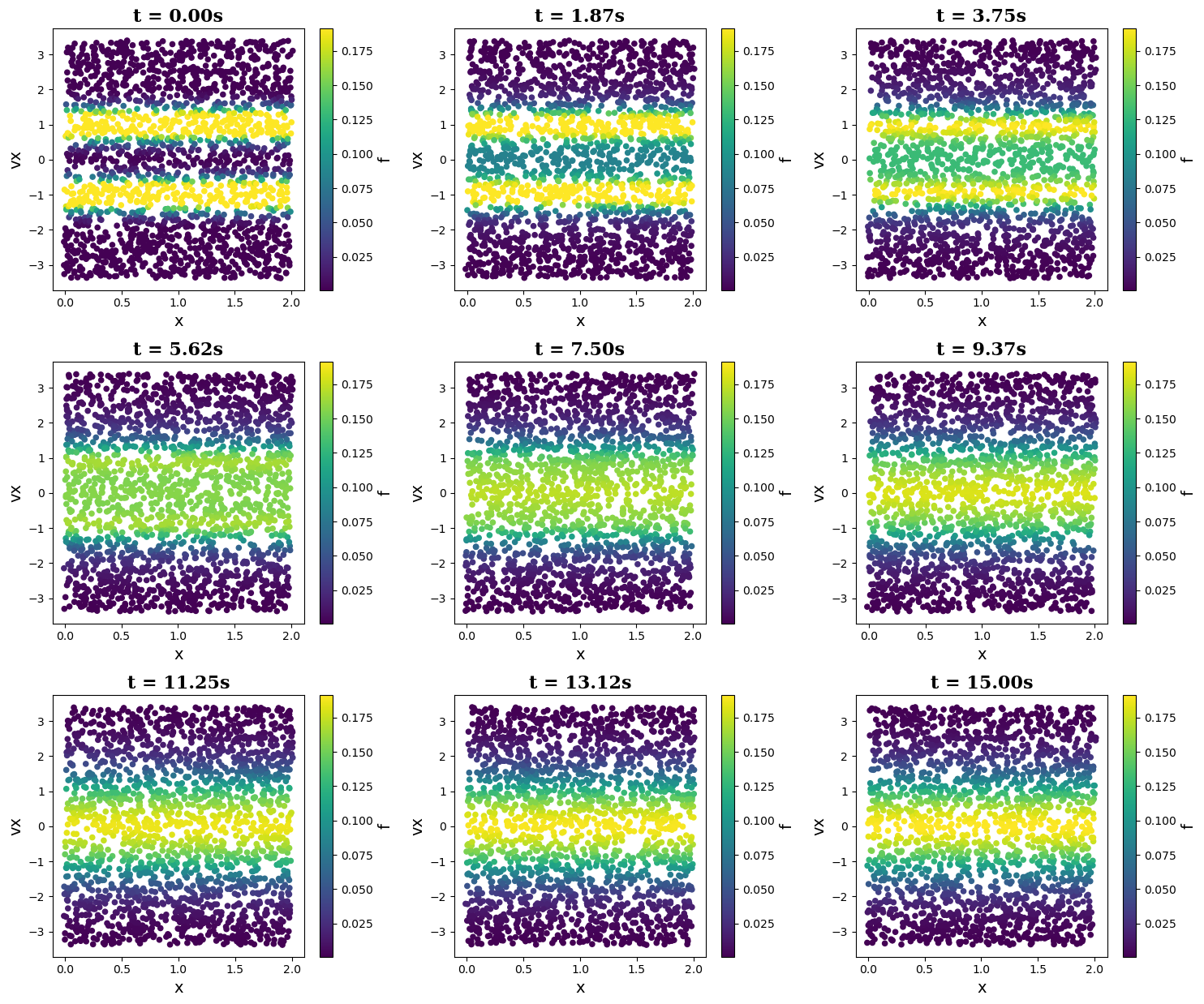}
\caption{Phase space evolution showing the relaxation to spatial and velocity homogeneity.}
\label{fig:collision_phase_space}
\end{subfigure}
\hfill
\begin{subfigure}{0.48\textwidth}
\includegraphics[width=\textwidth]{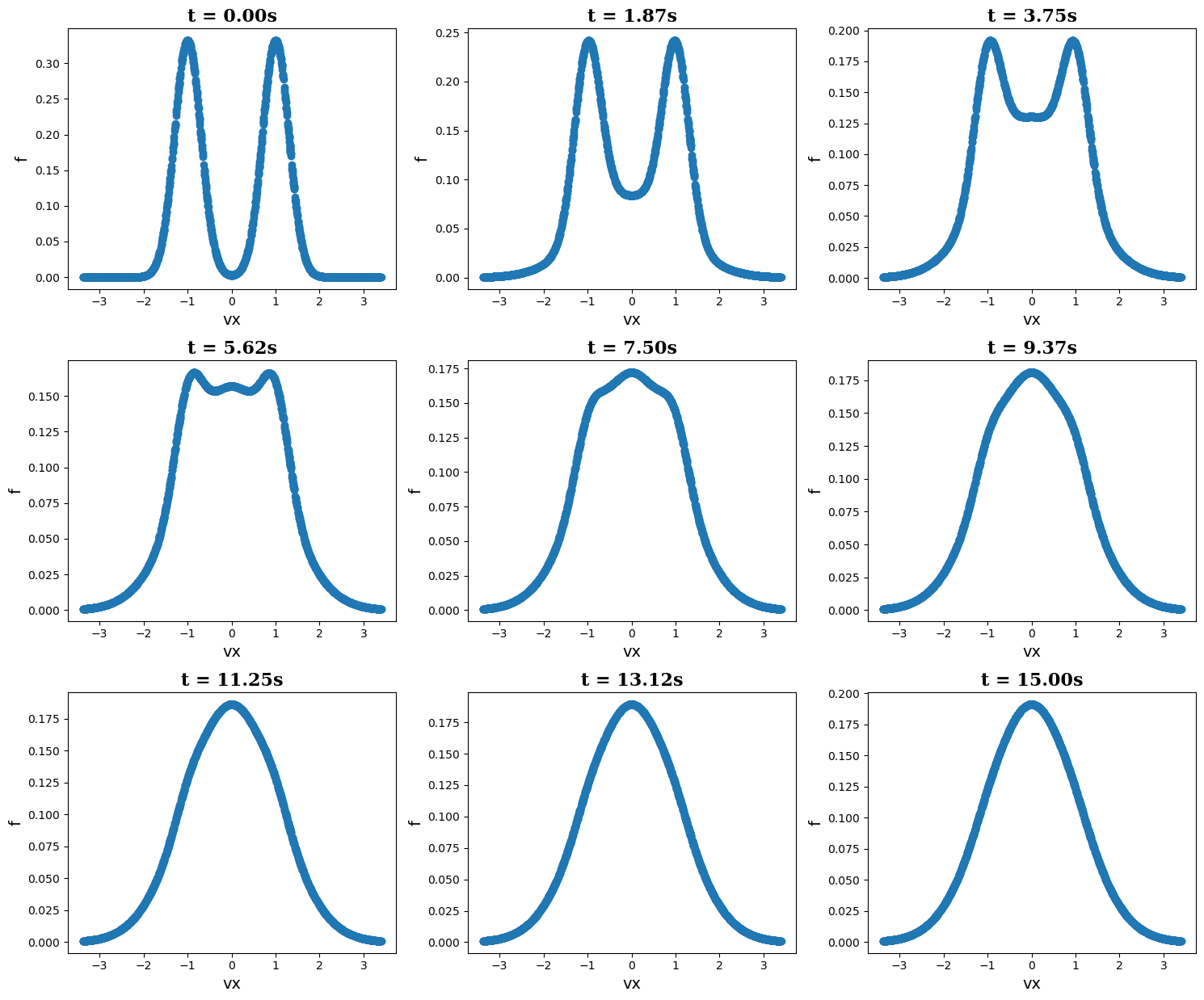}
\caption{Evolution of the velocity distribution showing merging of the two peaks.}
\label{fig:collision_velocity}
\end{subfigure}
\caption{Relaxation dynamics of the two-stream distribution under collisions.}
\label{fig:collision_evolution}
\end{figure}

Figure~\ref{fig:collision_evolution} illustrates the relaxation process. The phase space distribution (Fig.\ref{fig:collision_velocity}) shows the merging of the initial two-peak structure into a single Maxwellian.

\begin{figure}[H]
\centering
\includegraphics[width=0.8\textwidth]{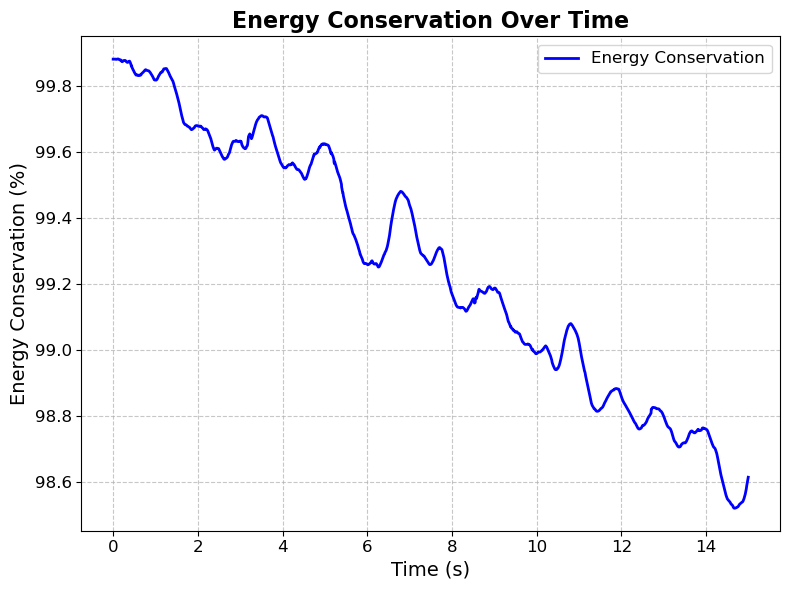}
\caption{Time evolution of the system's total thermal energy, showing conservation within 1.4\% over the simulation duration.}
\label{fig:energy_conservation}
\end{figure}

The evolution of total thermal energy (Fig.~\ref{fig:energy_conservation}) shows a small drift of approximately 1.4\% over 3,000 time steps. While this drift is acceptable for many applications, it could be further reduced through the use of implicit time integration schemes or smaller time steps. Notably, this energy drift is specific to the collisional case, as our previous collisionless simulations exhibited good energy conservation.

These results demonstrate the capability of our method to handle both collisional and collisionless plasma dynamics within a unified framework, while maintaining good conservation properties and computational efficiency.

\section{Discussion on Computational Cost}

In this work, we have not provided a direct quantitative comparison of the computational runtime between the conventional Particle-in-Cell (PIC) approach and our flow-based method. Such a comparison is strongly dependent on implementation details, including programming language, data structure choices, and optimization strategies (e.g., compiler flags, specialized libraries, parallelization). Ensuring that both methods receive precisely the same level of code optimization is also challenging. Therefore, to avoid misleading conclusions, we only provide a theoretical cost analysis in this section.

\subsection{Costs in PIC and Flow-Based Methods}

In both approaches, the computational workload can be broken down into three primary steps:

\begin{enumerate}
    \item \textbf{Field computation and interpolation:} Obtain the electric (or electrostatic) field on the grid, and interpolate it to the particle (or marker) locations.
    \item \textbf{Particle/marker push:} Advance the particles (or markers) forward in time according to the equations of motion.
    \item \textbf{Collision or distribution update:} In PIC, perform Monte Carlo (MC) collisions; in the flow-based method, update the distribution function represented by the markers.
\end{enumerate}

We assume the same number of grid points, denoted by \(M\), and a time advancement scheme whose total number of iterations is proportional to the number of particles or markers. 

Let \(N_{\mathrm{PIC}}\) denote the number of PIC particles. Let \(N_{\mathrm{flow}}\) denote the number of markers in the flow-based method. Let \(N_{\mathrm{quad}}^{\mathrm{qs}}\) and \(N_{\mathrm{quad}}^{\mathrm{ff}}\) represent the number of Gaussian quadrature points employed in the quasi-static and full field solutions, respectively.

Table~\ref{tab:compare} summarizes the main theoretical costs associated with both methods per time step (excluding constants and lower-order terms). We assume a solver cost of \(\mathcal{O}(M \log M)\) for the fields in both approaches \cite{genovese2006efficient}.

\begin{table}[ht]
\centering
\caption{Theoretical cost comparison of PIC vs.\ flow-based method per time step.}
\label{tab:compare}
\begin{tabular}{lccc}
\toprule
\textbf{Operation} & \textbf{PIC} & \multicolumn{2}{c}{\textbf{Flow}} \\
\cmidrule(lr){3-4}
 & & \textbf{Quasi-static} & \textbf{Full field} \\
\midrule
Charge/Current deposition  & \(\mathcal{O}(N_{\mathrm{PIC}})\) & - & \(\mathcal{O}(M*N_{\mathrm{quad}}^{ff})\)\\
Field solve       & \(\mathcal{O}(M \log M)\)         &  \(\mathcal{O}({N_{\mathrm{quad}}^{qs}}* N_{\mathrm{flow}})\) & \(\mathcal{O}(M \log M)\)\\
Force interpolation                & \(\mathcal{O}(N_{\mathrm{PIC}})\) &  \multicolumn{2}{c}{\(\mathcal{O}(N_{\mathrm{flow}})\)}\\
Particle/marker push               & \(\mathcal{O}(N_{\mathrm{PIC}})\) &  \multicolumn{2}{c}{\(\mathcal{O}(N_{\mathrm{flow}})\)} \\
Collision/distribution update   & \(\mathcal{O}(N_{\mathrm{PIC}})\) & \multicolumn{2}{c}{\(\mathcal{O}(N_{\mathrm{flow}})\)} \\
\bottomrule
\end{tabular}
\end{table}

From Table~\ref{tab:compare}, one observes that the flow-based method can substantially reduce costs related to pushing and updating particles when \(N_{\mathrm{flow}} \ll N_{\mathrm{PIC}}\). However, the overhead for field-related operations is nontrivial, especially in higher-dimensional phase space where \(N_{\mathrm{quad}}\) can grow quickly. We will compare the computational cost differences in field computation in the subsection below.

\subsection{Implications in Practice}

While it is challenging to predict how the computational cost scales across different spatial and velocity dimensions, we can provide a rough estimation. Let \(d_{\mathrm{space}}\) be the number of spatial dimensions and \(d_{\mathrm{velo}}\) be the number of velocity dimensions, so that the total phase-space dimension is
\[
d = d_{\mathrm{space}} + d_{\mathrm{velo}}.
\]
Let \(grid_{1d}\) be the number of grid points in one spatial dimension, implying that the total number of grid cells scales as \(grid_{1d}^{\,d_{\mathrm{space}}}\). In a standard PIC simulation, if each grid cell has approximately \(k_{\mathrm{cell}}\) macroparticles, the total PIC particle count is
\[
N_{\mathrm{PIC}} \approx k_{\mathrm{cell}} \,\bigl(grid_{1d}^{\,d_{\mathrm{space}}}\bigr).
\]
(Strictly, one may need more particles per cell as \(d_{\mathrm{velo}}\) increases to maintain accuracy, but we keep \(k_{\mathrm{cell}}\) fixed here for simplicity.)

In the quasi-static field solution, quadrature $N_{\mathrm{quad}}^{qs}$ are required due to integration over the full space within a \emph{compact support}, typically scaling as $(q_{qs})^d$, where $q_{qs}$ denotes the number of quadrature points required in single dimension for integration within the compact support of one kernel.

In contrast, the full field solution involves a marginal integration over the velocity subspace, resulting in quadrature points $N_{\mathrm{quad}}^{ff}$ that scale as $(q_{ff})^{d_{velo}}$, where $q_{ff}$ denotes the number of quadrature points required in single dimension for marginal integration at one spacial location.

\paragraph{Example:}

In our 1D1V (\(d_{\mathrm{space}}=1,\,d_{\mathrm{velo}}=1\)) Landau damping test, we observed a marker-to-particle ratio of approximately \(1{:}100\). Specifically, with around \(350{,}000\) PIC particles and \(3{,}500\) flow markers, we achieve comparable accuracy. 

we assume that this relationship of accuracy scales (See section \ref{sec:Error Estimation}) with the integration accuracy with a constant factor:

\begin{subequations}
\begin{align}
\frac{k}{N_{\mathrm{flow}}^{\,1 + \tfrac{2}{d+1}}}&=\frac{1}{N_{\mathrm{PIC}}^{0.5}} \\
N_{\mathrm{flow}} 
\;& \approx\;
\bigl(kN_{\mathrm{PIC}}^{\,0.5} \bigr)^{\,\tfrac{d+1}{d+3}},
\end{align}
\end{subequations}

Some representative parameters from the 1d1v landau damping test are:
\[
k_{\mathrm{cell}} = 10935,\quad grid_{1d} = 32,\quad k \; = 1360
\]

We assume \(q_{qs} = 4\) and \(q_{ff} = 50\). Notice that \(q_{ff}\) is much larger than \(q_{qs}\) due to the difference in integration areas: in the quasi-static case, each integration is performed within the compact support of each kernel, whereas in the full-field solution, the integration extends over the entire velocity space.

Using these parameters, one can form approximate scalings to predict 
\[
\frac{N_{\mathrm{PIC}}}{N_{\mathrm{flow}}}
,\quad
\frac{N_{\mathrm{PIC}}}{q_{qs}^{\,d}\,\cdot\,N_{\mathrm{flow}}}
,\quad \text{and}\quad 
\frac{N_{\mathrm{PIC}}}{q_{ff}^{\,d_{velo}}\cdot M}.
\]
Table~\ref{tab:scaling_examples} shows indicative values for various combinations of \(d_{\mathrm{space}}\) and \(d_{\mathrm{velo}}\). 

Here, \(\tfrac{N_{\mathrm{PIC}}}{N_{\mathrm{flow}}}\) highlights how many fewer markers might be required compared to particles. The other two expressions capture the relative computational cost of field computation in PIC compared to quasi-static or full-field solutions. When interpreting these two ratios, it is more important to focus on how each quantity changes with dimension rather than to compare them directly. Although they both scale with the relative computational cost, the proportionality factors themselves can vary significantly.

\begin{table}[ht]
\centering
\caption{Approximate ratios for \(\tfrac{N_{\mathrm{PIC}}}{N_{\mathrm{flow}}}\) , \(\tfrac{N_{\mathrm{PIC}}}{q_{qs}^{\,d} \times N_{\mathrm{flow}}}\) and \(\tfrac{N_{\mathrm{PIC}}}{q_{ff}^{\,d_{velo}} \times M}\) under various dimensions, using parameters motivated by the 1D1V Landau damping case, indicating relative computational costs}
\label{tab:scaling_examples}
\begin{tabular}{ccc|ccc}
\toprule
\(\boldsymbol{d_{\mathrm{space}}}\) & \(\boldsymbol{d_{\mathrm{velo}}}\) & \(\boldsymbol{d}\) 
& \(\frac{N_{\mathrm{PIC}}}{N_{\mathrm{flow}}}\) & \(\frac{N_{\mathrm{PIC}}}{q_
{qs}^{\,d}\times N_{\mathrm{flow}}}\) &\(\frac{N_{\mathrm{PIC}}}{q_
{ff}^{\,d_{velo}}\times M}\) \\
\midrule
1 & 1 & 2 & 100    & 6.259 & 218.7\\
1 & 2 & 3 & 40.45  & 0.632 &4.375\\
1 & 3 & 4 & 21.17  & 0.0827 &0.08748\\
2 & 2 & 4 & 196.50 & 0.767 &4.375\\
2 & 3 & 5 & 113.60 & 0.110 &0.08748\\
3 & 3 & 6 & 617.20 & 0.150 &0.08748\\
\bottomrule
\end{tabular}
\end{table}

From Table~\ref{tab:scaling_examples}, the flow-based method can require fewer markers (sometimes by one or two orders of magnitude) compared to the number of particles used in PIC. However, its field computation is some time more expensive in higher dimensional phase space due to the quadrature factor \(q_{qs}^{\,d}\) or \(q_{ff}^{\,d_{velo}}\). Still, even with this overhead, the overall time of computation would be largely reduced using flow based method.

Of course, the actual performance also depends on implementation details and hardware. Nonetheless, these estimates suggest that for many multi-dimensional problems, the flow-based method can provide considerable savings in time-to-solution relative to PIC.

\section{Conclusions}
\label{sec:conclusions}

We have presented a unified computational framework for kinetic plasma simulations that combines core ideas from direct Vlasov solvers, particle-in-cell (PIC) schemes, and continuous normalizing flows (CNFs). By adopting the characteristic viewpoint of the Vlasov equation together with the instantaneous-change-of-variables perspective from normalizing flows, our method transports the distribution function \( f(\mathbf{z}, t) \) explicitly along phase-space trajectories. This strategy waives the necessity of high-dimensional grids (as in direct Vlasov solvers) and mitigates the sampling noise commonly observed in Monte Carlo-based PIC. Below, we summarize the main features and findings.

\subsection*{Methodological Innovations}

\begin{itemize}
\item We introduced a hybrid framework that bridges deterministic and particle-based approaches by implementing distribution-bearing markers that carry both phase-space coordinates and distribution function values. Numerical experiments demonstrated accuracy comparable to or exceeding traditional PIC methods while using orders of magnitude fewer markers.

\item A scatter-point integration scheme, inspired by SPH methods, was developed for electromagnetic field calculations. This approach replaces conventional grid-based operations with kernel-weighted integral evaluations, showing particular efficacy in reducing sampling noise for canonical plasma physics benchmarks such as Landau damping and two-stream instability.

\item The framework naturally accommodates collisional physics through direct modification of marker distribution values, as demonstrated using the BGK collision model. This formulation successfully captured collisional relaxation phenomena while maintaining reasonable energy conservation.
\end{itemize}

\subsection*{Technical Capabilities}

\begin{itemize}
\item The method supports adaptive refinement through dynamic marker insertion in regions of interest. A novel backward-in-time integration scheme enables accurate initialization of new markers, allowing targeted computational resource allocation.

\item Performance analysis revealed that despite additional integral evaluations, the reduction in required markers often leads to favorable computational scaling compared to PIC methods, particularly for problems involving fine-scale phase-space structures.
\end{itemize}

\subsection*{Limitations and future directions.}  
This paper focuses on the fundamental formulation and preliminary tests of the flow-based method. Below, we discuss potential extensions and highlight future research directions:

\begin{itemize}
    \item \textit{Choice of kernel and its impact on accuracy and physics.}  
    We employed a cubic spline kernel, commonly used in SPH. Other kernel functions may offer advantages in terms of accuracy and noise control, but their performance remains to be explored thoroughly.

    \item \textit{Extensions beyond the BGK collision operator.}  
    While the BGK operator is intuitive and simple, it overlooks certain physical effects such as collision-induced radiation. By introducing more advanced collision operators, one can incorporate additional physics. For example, a stochastic formulation,
    \begin{subequations}
    \label{eq:general_sde}
    \begin{align}
        d\mathbf{x} &= \mathbf{v}\,dt, \\
        d\mathbf{v} &= \frac{q}{m}\bigl(\mathbf{E} + \mathbf{v}\times\mathbf{B}\bigr)\,dt 
                      -\nu\mathbf{v}\,dt + \boldsymbol{\sigma}\,d\mathbf{W},
    \end{align}
    \end{subequations}
    transforms the deterministic flow into a stochastic differential equation (SDE). Correspondingly, the instantaneous change-of-variables formula must be updated for what are known as stochastic normalizing flows~\cite{wu2020stochastic, hodgkinson2021stochastic}.

    \item \textit{Improved time-integration schemes.}  
    We use a leapfrog integrator in this paper, which performed well in simpler test cases. For enhanced accuracy and better energy conservation---and to allow larger time steps---higher-order or implicit methods are preferred. Early experiments with a 4th-order Yoshida symplectic integrator~\cite{yoshida1990construction} are promising, and we anticipate that implicit formulations (e.g., incorporating gyrokinetic approximations) will be particularly important for simulations in high magnetic field regimes, where the electron plasma frequency may drastically limit time steps.
\end{itemize}

In summary, our numerical experiments indicate that a flow-based approach to kinetic plasma simulations offers notable flexibility, accuracy, and noise-reduction benefits over traditional techniques. By unifying deterministic Vlasov and particle-based elements, we have a framework capable of capturing high-dimensional phase-space dynamics at modest computational cost. We expect that further refinements, guided by practical plasma applications, will enhance both the capabilities and the relevance of this methodology for cutting-edge plasma physics research.

% \appendix
% \section{Example Appendix Section}

% \cite{lamport94}.
%%       \cite{<label>} ==> [1]

%%
\bibliographystyle{elsarticle-num}
\bibliography{main}  % references is the name of your .bib file without extension
% Example citation, See \cite{lamport94}.

% %% If you have bib database file and want bibtex to generate the
% %% bibitems, please use
% %%
% %%  \bibliographystyle{elsarticle-num} 
% %%  \bibliography{<your bibdatabase>}

% %% else use the following coding to input the bibitems directly in the
% %% TeX file.

% %% Refer following link for more details about bibliography and citations.
% %% https://en.wikibooks.org/wiki/LaTeX/Bibliography_Management


% %% For numbered reference style

%   \textit{\LaTeX: a document preparation system},
%   Addison Wesley, Massachusetts,

% \end{thebibliography}
\end{document}